\documentclass[twocolumn,aps,prb,superscriptaddress,showpacs]{revtex4}
\usepackage{graphicx}
\usepackage{amsmath}
\usepackage{amsfonts}

\begin{document}


\title{Interaction Driven Quantum Phase Transitions in Fractional Topological Insulators}

\author{Hua Chen}
\affiliation{Zhejiang Institute of Modern Physics, Zhejiang University, Hangzhou 310027, P.R. China}
\affiliation{National High Magnetic Field Laboratory and Department of Physics, Florida State University, Tallahassee, Florida 32306, USA}

\author{Kun Yang}
\affiliation{National High Magnetic Field Laboratory and Department of Physics, Florida State University, Tallahassee, Florida 32306, USA}

\date{\today}

\begin{abstract}
We study two species of (or spin-1/2) fermions with short-range intra-species repulsion in the presence of opposite (effective) magnetic field, each at Landau level filling factor 1/3. In the absence of inter-species interaction, the ground state is simply two copies of the 1/3 Laughlin state, with opposite chirality, representing the fractional topological insulator (FTI) phase. We show this phase is stable against moderate inter-species interactions. However strong enough inter-species repulsion leads to phase separation, while strong enough inter-species attraction drives the system into a superfluid phase. We obtain the phase diagram through exact diagonalization calculations. The FTI-superfluid phase transition is shown to be in the (2+1)D XY universality class, using an appropriate Chern-Simons-Ginsburg-Landau effective field theory.
\end{abstract}

\maketitle

\section{Introduction}

Topological phases of matter are of strong interest to physicists\cite{wenbook}. Exciting recent developments in this area are the discoveries of topological insulators, which can be viewed as time-reversal invariant analogs of integer quantum Hall (IQH) states\cite{reviews}. A natural question is if the nature supports time-reversal invariant analogs of {\em fractional} quantum Hall (FQH) states, namely fractional topological insulators (FTIs). Just like FQH states, FTIs are expected to be stabilized by strong interaction.

Perhaps the simplest FTI state\cite{bernevig06} is two (decoupled) copies of Laughlin states for up- and down-spin electrons, with {\em opposite} chirality. Such states are the exact ground states of model Hamiltonians\cite{haldane83} with special short-range repulsions between electrons with the same spin, but no interaction between electrons with opposite spins. More recently lattice models that support FTIs have been constructed in both two dimension (2D)\cite{neupert} and three dimension (3D)\cite{levin11}. Since FTIs are stabilized by specific forms of interactions, it is important to understand the stability of FTIs when interactions are varied, and in particular, what kind of competing phases FTIs yield to when quantum phase transitions (QPTs) are triggered by varying interactions. Also of strong interest is the nature of such QPTs, and the critical properties of the QPT when it is continuous.

Motivated by the above we consider the stability of the simplest FTI state\cite{bernevig06} in the presence of more general electron-electron interaction, in particular inter-spin interactions. We show that sufficiently strong inter-spin repulsion leads to phase separation, while inter-spin attraction drives the system into a superfluid (SF) phase. The FTI-SF transition is 2nd order, and in the 3D XY universality class. We also show that the FTI and SF phases have natural analogs in bilayer quantum Hall systems at total filling factor 1 with layer imbalance\cite{Champagne}; our results are thus relevant to that system which is of strong interest in its own right.

The rest of the paper is organized as follows. In Sec.~\ref{model} we introduce the simplified model and discuss the related symmetry. In Sec.~\ref{phase} we focus on the numerical result on various properties of different phases, including energy spectrum, pair correlation function, and global phase diagram. The corresponding effective field theory is constructed in Sec.~\ref{effective}. Finally we discuss the essential relation between SF phase and bilayer layer quantum Hall systems at total filling factor 1 in Sec.~\ref{bilayer}.

\section{Model and Symmetry}
\label{model}

In this work we use a simple model to study FTIs, in which Landau levels are created by the spin-orbit coupling in the presence of a strain\cite{bernevig06,Ghaemi}. Unlike external magnetic field, strain does {\em not} break  time reversal symmetry. As a result electrons with opposite spins experience opposite {\em effective} orbital magnetic fields, and Landau levels with opposite chirality appear for opposite spin orientations. The single-particle Hamiltonian is given by $\hat{\mathcal{H}}^{\sigma}_0(\mathbf{r})=\frac{1}{2m}(\hat{\mathbf{p}}+\frac{e}{c}\mathbf{A}^{\sigma})^2$ with spin dependent vector potential $
\mathbf{A}^\sigma=\sigma B_{\text{eff}}(y,0,0)$, with $\sigma=\pm 1$ for up- and down-spin electrons.
Ref. \onlinecite{bernevig06} discussed how to realize such a situation in specific semiconductor materials; here we use this as an idealized model to study the stability of FTIs.

Just like in FQH states electrons are confined to the lowest Landau level, and electron-electron interaction dominates. In this work, we consider {\em hard-core} interaction $V_1 l_\text{B}^4\nabla^2\delta^{(2)}(\mathbf{r})$ between same spin fermions, and $V_0 l_\text{B}^2\delta^{(2)}(\mathbf{r})$ between opposite spin fermions; $l_B$ is magnetic length. The intraspin interaction $V_1$ is the energy of a pair of electrons with relative angular momentum $1$ which is first introduced by Haldane\cite{haldane83}. It captures the essence of the topological phases and gives rise to the exact model ground state, Laughlin state. The interacting Hamiltonian $\hat{\mathcal{H}}_{int}$ is defined by setting $V_1=V\sin(\varphi)$ and $V_0=V\cos(\varphi)$, where the parameter $\varphi$ tunes the relative strength between the $V_1$ and $V_0$ terms. Throughout the paper we set the overall energy scale $V$ as energy unit and magnetic length $l_\text{B}$ as length unit. We also use torus geometry by imposing (magnetic) periodic boundary conditions. In second quantization the interacting Hamiltonian reads
\begin{eqnarray}
\label{Hamint}
 \hat{\mathcal{H}}_{int} &=&
  \frac{1}{2}\sin(\varphi)\sum_{\{j_i\}\sigma}V_{j_1j_2j_3j_4}^{\sigma \sigma}
   c_{j_1\sigma}^{\dagger}c_{j_2\sigma}^{\dagger}c_{j_3\sigma}c_{j_4\sigma} \nonumber\\
  &+&\frac{1}{2}\cos(\varphi)\sum_{\{j_i\}\sigma}V_{j_1j_2j_3j_4}^{\sigma \overline{\sigma}}
   c_{j_1\sigma}^{\dagger}c_{j_2\overline{\sigma}}^{\dagger}c_{j_3\overline{\sigma}}c_{j_4\sigma},
\end{eqnarray}
where $c^{\dagger}_{j\sigma}$ is the creation operator for an electron with spin index $\sigma$ and Landau orbital index $j$.

The full symmetry analysis of FQH systems at {\em rational} filling factors was provided by Haldane~\cite{haldane85}. Here we generalize his analysis, and show eigen states of our system can be characterized by a 2D wave vector $\mathbf{K}$ for {\em arbitrary} filling factor. To prove this we introduce magnetic translation operator for particle $j$ with spin $\sigma$: $\hat{T}_{j}^{\sigma}(\mathbf{r})=\exp\left\{ i\mathbf{r}\cdot\mathbf{\Pi}_{j}^{\sigma}/\hbar\right\}$, where $\mathbf{\Pi}_{j}^{\sigma}=\hat{\mathbf{p}}_j+\frac{e}{c}\mathbf{A}^{\sigma}({\bf r}_j)$. $\hat{T}_{j}^{\sigma}(\mathbf{r})$ commutes with $\hat{\mathcal{H}}_{0}^{\sigma}(\mathbf{r}_{j})$, but $\hat{T}_{j}^{\sigma}(\mathbf{a})\hat{T}_{j}^{\sigma}(\mathbf{b})=\exp\left\{ -i\sigma\hat{\mathbf{z}}\cdot(\mathbf{a}\times\mathbf{b})\right\} \hat{T}_{j}^{\sigma}(\mathbf{b})\hat{T}_{j}^{\sigma}(\mathbf{a})$ and thus do {\em not} commute with each other in general, due to the Berry phase induced by magnetic field. We now introduce the center-of-mass translation operator: $\hat{T}\left(\mathbf{r}\right)=\prod_{j,\sigma}\hat{T}_{j}^{\sigma}\left(\mathbf{r}\right)$. Let $\hat{T}_{X}$ and $\hat{T}_{Y}$ be center of mass translations by some arbitrary amounts along $x$ and $y$ directions. Both $\hat{T}_{X}$ and $\hat{T}_{Y}$ commute with the many body Hamiltonian $\hat{\mathcal{H}}$ which can be separated into a center of mass term and relative motion terms. Since {\em opposite orientation} spins experience {\em opposite} orbital magnetic fields in the present system, the Berry phases picked up by up- and down-spin electrons cancel when they have the same numbers, as a result of which $\hat{T}_{X}\hat{T}_{Y}=\hat{T}_{Y}\hat{T}_{X}$. Thus the Hamiltonian can be {\em simultaneously} diagonalized along with $\hat{T}_{X}$ and $\hat{T}_{Y}$, with eigen states labeled by a 2D momentum quantum number.

\begin{figure} [htp]
\includegraphics[width=0.48\textwidth]{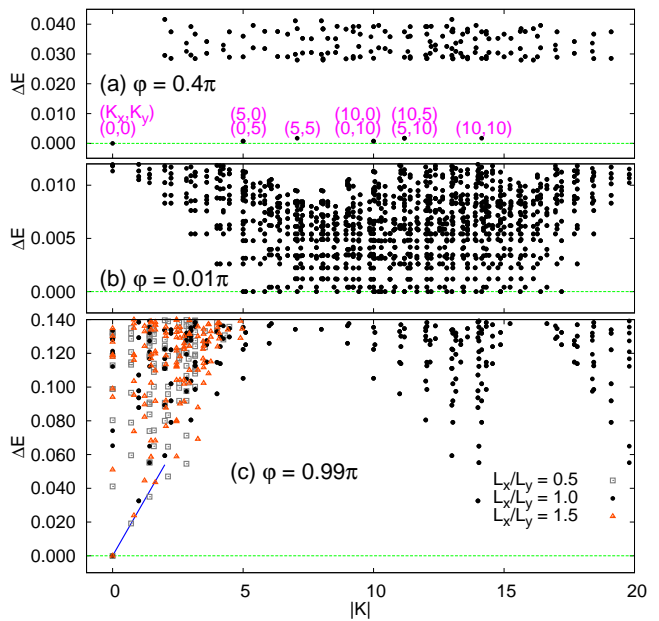}
\caption{\label{fig:energy} (Color online) Representative low-energy excitation spectra, obtained by exact diagonalization studies on a torus at $\nu_\uparrow=\nu_\downarrow=1/3$ with number of flux quanta $N_{f}=15$, and $N_\uparrow=N_\downarrow=5$, mostly based on square geometry with magnetic periodic boundary conditions. The states are labeled by the magnitude of the conserved wavevector $|\mathbf{K}|$ in units of $2\pi/\sqrt{L_{x}L_{y}}$. (a) The fractional topological insulator at $\varphi = 0.4\pi$. Wave vectors of 9 nearly degenerate ground states are marked. They are  separated from excited states by a clear gap. (b) Strong inter-spin repulsion at $\varphi = 0.01\pi$ that leads to phase separation between up- and down-spin electrons. (c) Strong inter-spin attraction at $\varphi = 0.99\pi$ that stabilizes a paired superfluid phase. The {\em unique} ground state is at $\mathbf{K}=\mathbf{0}$, and low-energy excitations form a Goldstone mode with linear dispersion. Data from systems with the same size but different rectangular geometries are included in (c) to obtain more $\mathbf{K}$'s in order to reveal the linear dispersion more clearly.}
\end{figure}

\section{Phase diagram}
\label{phase}
The simplest FTI states, in our language, corresponds to the ground states at $\varphi_\text{F}=\pi/2$, in which case $V_0=0$ and thus the two spin species decouple, and there are $3\times 3=9$ {\em exactly} degenerate ground states that are separated from excited states by a finite gap. Such degeneracy is a {\em topological} property of the phase, and robust against small perturbations\cite{wenbook}.

As shown in FIG.~\ref{fig:energy} (a), when we turn on a small $V_0$ by having $\varphi=0.4\pi$, we still have 9 nearly degenerate low lying states that are well separated from excited states by a gap; this clearly indicates the system is still in the FTI phase.

\begin{figure} [htp]
\includegraphics[width=0.48\textwidth]{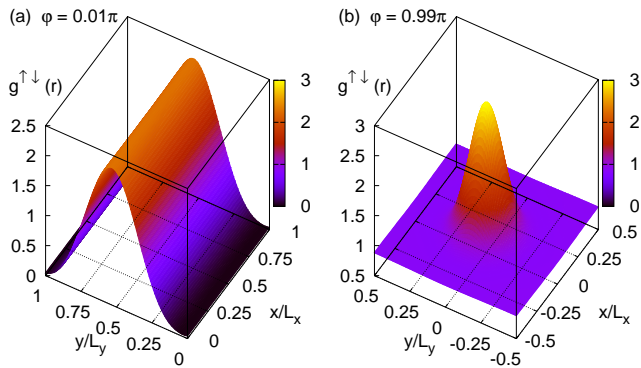}
\caption{\label{fig:pair} (Color online) Two-particle correlation function $g^{\uparrow\downarrow}(\mathbf{r})$ between different spin orientations, obtained by exact diagonalization on a torus with square geometry $L_x=L_y$ for number of electrons $N^{\uparrow}=N^{\downarrow}=5$ and number of flux quanta $N_f=15$, as a function of ($x/L_{x}$,$y/L_{y}$). (a) The large hump around $y/L_{y}=0.5$ with $\mathbf{K}=(5,7)$ indicates the phase separation\cite{symmetry}. (b) The maximum of $g^{\uparrow\downarrow}$ at the origin is indicative of the paired character of the ground state with $\mathbf{K}=\mathbf{0}$ which has highest symmetry in the Brillouin zone.}
\end{figure}

However sufficiently strong $V_0$ {\em destabilizes} the FTI phase.
As shown in FIG.~\ref{fig:energy} (b), at $\varphi=0.01\pi$, where $V_0\gg V_1>0$, there are many low-lying states with {\em no} gap. In this case we expect the up- and down-spin electrons phase separate into $\nu_\uparrow=2/3, \nu_\downarrow=0$, and $\nu_\downarrow=2/3, \nu_\uparrow=0$ regions. This is supported by the opposite-spin pair correlation function $g^{\uparrow\downarrow}(\mathbf{r})=L_xL_y/(N^{\uparrow}N^{\downarrow}l_\text{B}^2)\langle\text{GS}| \sum_{ij}\delta^{(2)}(\mathbf{r}+\mathbf{r}_i^{\uparrow}-\mathbf{r}_j^{\downarrow})|\text{GS}\rangle$, as illustrated in Fig.~\ref{fig:pair} (a): It goes to zero at origin $\mathbf{r}=\mathbf{0}$, while shows a large hump around $y/L_y=0.5$, the farthest possible place in our finite size system (with periodic boundary condition). The low-lying states correspond to fluctuations of the boundary between the separated regions, which do not have clear patterns in their quantum numbers.

\begin{figure} [htp]
\includegraphics[width=0.48\textwidth]{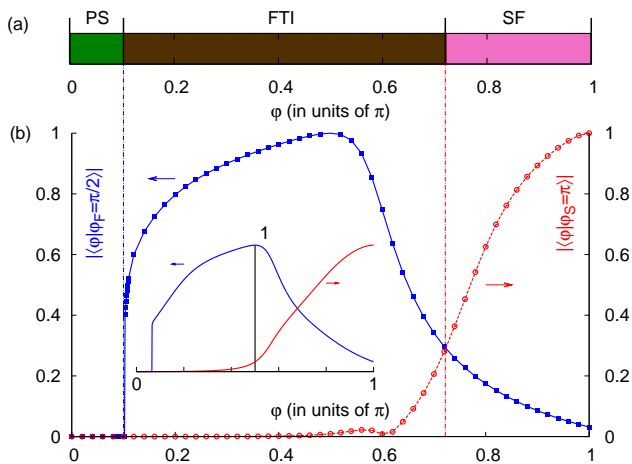}
\caption{\label{fig:overpd} (Color online) Overlaps between actual ground state with ideal model states for the fractional topological insulator ($\varphi_{\text{F}}=\pi/2$) and superfluid ($\varphi_{\text{S}}=\pi$) phases (b), and resultant phase diagram and boundaries (a). FTI, SF and PS stand for fractional topological insulator, superfluid and phase separation respectively. These overlaps are obtained by exact diagonalization on a torus with square geometry $L_x=L_y$ for number of electrons $N^{\uparrow}=N^{\downarrow}=5$ and number of flux quanta $N_f=15$. The inset presents results with system size $N_f=12$ (with $N^{\uparrow}=N^{\downarrow}=4$), showing very little size dependence.}
\end{figure}

The situation is more  interesting when we switch on {\em attractive} inter-species interaction, which can lead to pairing between opposite spin electrons. Such pairs see {\em zero} magnetic field, and can thus condense; their condensation leads to a superfluid (SF) phase with a spontaneously broken U(1) symmetry. This is what happens at $\varphi=0.99\pi$, where $-V_0\gg V_1>0$. The pairing character of ground state is clearly manifested in the pair correlation function $g^{\uparrow \downarrow}(\mathbf{r})$ in Fig.~\ref{fig:pair} (b), which sharply peaks at origin. Stronger evidence for the SF nature of the phase is in the spectrum of Fig.~\ref{fig:energy} (c), which  features a {\em non-degenerate} ground state at $\mathbf{K}=\mathbf{0}$, and a low-energy mode with linear dispersion at long wavelength. This linear mode is the SF Goldstone mode.

To determine the phase diagram of the system and the precise locations of the phase boundaries, we calculate overlaps of the actual ground state with ideal model states for the FTI and SF phases. The former is simply the direct product of two decoupled Laughlin states obtained at $\varphi_\text{F}=\pi/2$, and the latter is chosen for the ground state at $\varphi_\text{S}=\pi$, whose explicit wave function is Eq. (2) of Ref. \onlinecite{kun01} (more on this later). The results are presented in Fig.~\ref{fig:overpd}, from which it is clear that phase separation (PS) is a first-order instability of the FTI phase, while the FTI-SF transition appears to be continuous.

\section{Effective Field Theory}
\label{effective}
In this section we develop an effective field theory that describes the FTI and SF phases on equal footing, and in particular, the phase transition between them. Our starting point is the Chern-Simons-Landau-Ginzburg (CSLG) theory~\cite{cslg} for FQH states, in which the Laughlin state is viewed as a superfluid state of composite bosons made of electrons bound to appropriate amount of flux. Since the FTI can be viewed as two copies of the Laughlin state with {\em opposite} chirality, it is thus natural to attach {\em opposite} flux to electrons with opposite spins. This leads to the following effective field theory lagrangian density:
\begin{eqnarray}
L&=&L_\uparrow+L_\downarrow+L_{\text{p}}+L_{\text{cs}}
+g(\overline{\phi}\psi_\uparrow\psi_\downarrow+\phi\overline{\psi}_\uparrow\overline{\psi}_\downarrow),\label{L}\\
L_\sigma&=&\overline{\psi}_\sigma(i\partial_t-a_0^\sigma)\psi_\sigma-\mu|\psi_\sigma|^2\nonumber\\
&+&{1\over 2m}|(-i \nabla-{\bf A}^\sigma-{\bf a}^\sigma)\psi_\sigma|^2+\cdots,\label{La}\\
L_{\text p}&=&\overline{\phi}(i\partial_t-a_0^\uparrow-a_0^\downarrow)\phi-(2\mu-\delta)|\phi|^2\nonumber\\
&+&{1\over 2M}|(-i\nabla-{\bf a}^\uparrow-{\bf a}^\downarrow)\phi|^2+\cdots,\label{Lm}\\
L_{\text{cs}}&=&L^\uparrow_{\text{cs}}-L^\downarrow_{\text{cs}}={1\over 4\pi}{\pi\over \theta}\epsilon^{\mu\nu\lambda}\left[a^\uparrow_\mu\partial_\nu a^\uparrow_\lambda-a^\downarrow_\mu\partial_\nu a^\downarrow_\lambda\right].
\label{Lcs}
\end{eqnarray}
Here $\psi_\sigma$ are composite boson fields for (bosonized) electrons, while ${\bf a}^\sigma$ is the Chern-Simons (CS) gauge field that attaches flux to spin-$\sigma$ composite bosons, which cancels extenal gauge field ${\bf A}^\uparrow=-{\bf A}^\downarrow$ {\em in average}. $\theta$ determines the amount of flux attached to each  particle; in the present case $\theta=3\pi$ for the 1/3 Laughlin state. Notice the minus sign between $L^\uparrow_{\text{cs}}$ and $L^\downarrow_{\text{cs}}$; this indicates the flux is in {\em opposite directions} for up- and down-spin composite bosons. The presence of pairing (or attractive) interaction between up- and down-spin electrons is encoded by introducing a pair field $\phi$, with lagrangian density $L_p$. It represents bound state of up- and down-spin composite bosons, and sees {\em no} external gauge field ${\bf A}$, due to the cancelation between ${\bf A}^\uparrow$ and ${\bf A}^\downarrow$. On the other hand it couples equally to the CS gauge fields ${\bf a}^\uparrow$ and ${\bf a}^\downarrow$. Using terminology familiar in cold atom contexts, we call the parameter $\delta$ detuning; positive $\delta$ favors unbound electrons while negative $\delta$ favors pair formation. Thus decreasing $\delta$ corresponds to increasing pairing strength. The $g$ term in Eq. (\ref{L}) describes pair formation and decay processes. Generic density-density interactions among the particles are kept implicit and represented by $\cdots$. Due to the relation between particle and CS flux density enforced by the CS terms, such density-density interactions can also be written in terms of CS flux density $b^\sigma=\epsilon^{ij}\partial_i a^\sigma_j$.

The distinction between the FTI and SF phases is the following. In the FTI phase, which is topologically equivalent to two independent FQH states, composite bosons with up- and down-spins both condense, namely $\langle\psi_\sigma\rangle\ne 0$\cite{cslg}; this gives rise to Anderson-Higgs masses to the CS gauge fields, and the corresponding Meisnner effect is equivalent to incompressibility responsible for FQHE. Due to the $g$ term in Eq. (\ref{L}), this implies $\langle\phi\rangle\ne 0$. In the SF phase however, there is only a pairing gap that penalizes {\em imbalance} between up- and down-spin electron numbers, but {\em no} overall charge incompressibility. This suggests in the SF phase $\langle\psi_\sigma\rangle = 0$ while we {\em still} have $\langle\phi\rangle\ne 0$, as $\phi$ couples {\em only} to the combination ${\bf a}^\uparrow+{\bf a}^\downarrow$ whose flux is the {\em difference} between up- and down-spin electron density. Thus the FTI and SF phases differ by one U(1) condensate, and the transition is driven by the appearance/disappearance of this condensate. The situation is somewhat similar to a transition between integer and fractional quantum Hall states driven by pairing interaction studied in Ref. \onlinecite{qhtranskun}. In particular, if we completely suppress the fluctuations of CS gauge fields, the critical theory will reduce to that of the (2+1)D XY model just as in Ref. \onlinecite{qhtranskun}, with the critical U(1) field being
$\psi_-=(\psi_\uparrow-\overline{\psi}_\downarrow)/\sqrt{2}$.

Also like in Ref. \onlinecite{qhtranskun}, fluctuations of CS gauge fields changes the physics, but in a very different manner here.
To proceed, we introduce new combinations of CS gauge fields:
$a_\mu^{\pm}=(a_\mu^\uparrow\pm a_\mu^\downarrow)/2$,
in terms of which Eqs. (\ref{La},\ref{Lm},\ref{Lcs}) take the form
\begin{eqnarray}
\label{Lup}
L_\uparrow&=&\overline{\psi}_\uparrow(i\partial_t-a_0^+-a_0^-)\psi_\uparrow-\mu|\psi_\uparrow|^2\nonumber\\
&+&{1\over 2m}|(-i\nabla-{\bf A}-{\bf a}^+-{\bf a}^-)\psi_\uparrow|^2+\cdots,\\
\label{Ldown}
L_\downarrow&=&\overline{\psi}_\downarrow(i\partial_t-a_0^++a_0^-)\psi_\downarrow-\mu|\psi_\downarrow|^2\nonumber\\
&+&{1\over 2m}|(-i\nabla+{\bf A}-{\bf a}^++{\bf a}^-)\psi_\downarrow|^2+\cdots,\\
L_{\text{p}}&=&\overline{\phi}(i\partial_t-2a_0^+)\phi-(2\mu-\delta)|\phi|^2\nonumber\\
&+&{1\over 2M}|(-i\nabla-2{\bf a}^+)\phi|^2+\cdots,\\
L_{\text{cs}}&=&{1\over \theta}\epsilon^{\mu\nu\lambda}a^+_\mu\partial_\nu a^-_\lambda.
\end{eqnarray}
Notice in particular the CS term is a bilinear coupling between $a^+$ and $a^-$, of the BF form\cite{bf}. Noting that since $\langle\phi\rangle\ne 0$, and thus $a^+$ acquires an Anderson-Higgs mass of the form $\Lambda a^{+j}a^+_j$ {\em throughout the phase diagram}, we can safely integrate $a^+_j$ out; due to their coupling to $a^-_j$ through $L_{\text{cs}}$, this will generate Maxwell terms of the form $F^{-0j}F^-_{0j}=e^{-j}e^-_j$. Integrating over $a^+_0$ enforces the proportionality constraint between $b^-=F^-_{ij}=\epsilon_{ij}\partial_i a^-_j$ and total density; this allows us to write some of the density-density interactions in ``$\cdots$" in the form of the remaining Maxwell terms $F^{-ij}F^-_{ij}=|b^-|^2$. Thus the effective action of $a^-$ after integrating out $a^+$ is that of Maxwell theory $F^{-\mu\nu}F^-_{\mu\nu}$, with space-time re-scaled according to the {\em non-universal} photon velocity, which is the Goldstone mode (or sound) velocity in the superfluid phase.

As a result the final critical theory for the FTI-SF transition is that of Ginzburg-Landau-Maxwell theory in (2+1)D, with Euclidean lagrangian density (after proper re-scaling of the critical field $\psi_-$), which is {\em dual} to the (2+1)D XY model\cite{dasguptahalperin}:
\begin{eqnarray}
\label{Leff}
L_{\text{eff}}[\psi_-, a^-]&=&|(\partial_\mu-ia_\mu^-)\psi_-|^2
-\lambda|\psi_-|^2+U|\psi_-|^4\nonumber\\
&+&\alpha F^{-\mu\nu}F^-_{\mu\nu}+\cdots.
\end{eqnarray}
Here ``$\cdots$" represent irrelevant terms involving higher order of derivatives or powers of fields (either in the original lagrangian or generated by integrating out $a^+$).
In the SF phase, $\langle\psi_-\rangle=0$, the $a^-$ gauge field is in the Coulomb phase which supports a {\em single} branch gapless photon mode with linear dispersion; this corresponds to the Goldstone mode of the SF phase. In the FTI phase $\langle\psi_-\rangle\ne 0$ (as a result we have two condensates, including $\langle\psi_+\rangle\ne 0$), and the $a^-$ gauge field is in the Higgs phase where there is {\em no} gapless excitation. The transition is in the (2+1)D XY universality class due to the duality.

\section{Relation with bilayer quantum Hall sysmtem at total filling factor $\nu_T=1$}
\label{bilayer}
Our model has close relation with bilayer quantum Hall system at total filling factor $\nu_T=\nu_{top}+\nu_{bottom}=1$ (where $\nu_{top}$ and $\nu_{bottom}$ correspond to filling factors of top and bottom layers respectively), especially in the presence of layer imbalance. This relation is revealed by making a particle-hole transformation\cite{kun01} for the lower layer and describe it in terms of {\em holes}, which results in hole filling factor $\nu_{hole}=1-\nu_{bottom}=\nu_{top}=\nu_{electron}$. Since the holes carry opposite charge as the electrons, the chirality due to external magnetic field is {\em opposite} for them as well. The ground state in the strong particle-hole pairing limit is nothing but the Halperin 111 state~\cite{halperin83} written in terms of electron coordinates, which is the ideal model state for the pseudospin ferromagnet\cite{yangbilayer} or interlayer exciton condensate state\cite{wenzee92}. The corresponding wave function written in terms of electrons in top layer and holes in bottom layer is presented in Ref. \onlinecite{kun01}. The specific case of $\nu_\uparrow=\nu_\downarrow=1/3$ correspond to $\nu_{top}=1/3$ and $\nu_{bottom}=2/3$, and the FTI and SF states discussed here correspond to decoupled single layer FQH and interlayer exciton condensate states. A transition between them has been observed experimentally\cite{Champagne}, and the theory developed here for the FTI-SF transition applies to that case as well. We note while bilayer quantum Hall liquid is a fascinating system in its own right, it may now also be used as testing ground for some of the theoretical predictions\cite{levinstern} and proposed techniques\cite{beri} to probe FTIs due to their close relation, given the fact that there is no candidate system for FTI at this point.

\section*{Acknowledgments}
This work was supported China Scholarship Council (HC) and NSF grant No. DMR-1004545 (KY).


\begin{thebibliography}{99}
\bibitem{symmetry} As required by time reversal symmetry, the reflection symmetry between the degenerate ground states with $\mathbf{K}=(5,7),(5,8),(10,7),(10,8)$ in phase separation region is observed in our numerical calculations.
\end{thebibliography}
\end{document}